# Internet of Things-based innovations in Saudi healthcare sector: A methodological approach for investigating adoption issues

**Feisal Hadi Masmali**
Newcastle Business School, The University of Newcastle, New South Wales, Australia
College of Business Administartion, Jazan University, Saudi Arabia
fmasmali@jazanu.edu.sa

**Shah Jahan Miah**
Newcastle Business School, The University of Newcastle, Newcastle City Campus, New South Wales, Australia
shah.miah@newcastle.edu.au

**Nada Yahya Mathkoor**
Engineering School, The University of Melbourne, Melbourne, Australia
nmathkoor@student.unimelb.edu.au

*Abstract* — *The Internet of Things (IoT) has proliferated over the last few years as the next-generation technologies that impact both human systems and businesses. Using today's Internet network capacities, this technology has extended various benefits in healthcare sectors. For instance, existing studies already indicated that information technology (IT) applications with IoT-based innovations may revolutionize the healthcare industry and subsequently help to improve the real-time reporting of patients' health data. It should be noted that the adoption of IoT and its relevant interventions in the health sector has not been as fast as the uptake been observed in other industries. To tackle this issue, we develop a qualitative phenomenological approach for investigating factors that affect IoT adoption and its integration into healthcare service delivery in Saudi Arabia. To this end, the study aims to reveal the practical experiences of healthcare professionals or service administrators in using IoT devices as they deliver medical services.*

## I INTRODUCTION

Internet of Things (IoT) is a term that Kevin Ashton coined in 1999 to describe a system in which the *"internet is used to empower computers to empower the world for themselves"* [1]. This view is associated to machines, infrastructure and humans through cloud computing [1]. Researchers such as, Breivold and Sandström [2] stated in their research that IoT helps connect every day apparatuses and devices that we use at home as well as industrial equipment. As many organizations as well as governments have realised the value of using IoT in their practices, the present study is intended to investigating factors that affect IoT adoption and its integration into healthcare service delivery in Saudi Arabia.

IoT research have significant association to e-government studies. E-government literature has grown rapidly over the past few decades in publication administration [3-5], agricultural service [6, 7], transportation [8, 9] and healthcare service [10-12]. However, e-government application in the healthcare sector have been limited and especially so for developing nations, due to many issues such as lack of investment and proper awareness of using the latest technologies for improving people's performance. This research makes an important contribution to the current e-government literature by creating a new direction of using IoT for enhancing e-government services specifically for healthcare, which involves adopting other latest mobile technologies. The system responds to the demand to understand the contexts in which e-government is implemented and especially with reference to IoT [13]; Shareef et al., 2016; Venkatesh et al., 2016; Wirtz & Daiser, 2018). The system will be very useful for establishing a better comprehension of using IoT in healthcare systems in other developing countries.

This paper extends TAM so that the adoption of IoT is better understood by examining the role of contextual factors in improving IoT in healthcare services. Current TAM-based IoT models concentrate mainly on human variables and devices' characteristics. This research illustrates the applicability of the expanded system to investigate IoT in the healthcare sector with empirical evidence in the developing countries' context. This study offers important research results for e-government improvements in Saudi Arabia. The significance of this study for e-government developers lies in providing them with valuable knowledge so that e-government healthcare strategies are effective and practical. This study will be conducted within the domain of Saudi's public healthcare sector. Under this public health care sector there are three different areas that we identified. One area is government public health which is the main channel of service delivery for the benefit of the rural and urban citizens. It is related to provide healthcare services ensuring community health support, advice and improvement for citizens.

## II THEORETICAL PERSPECTIVES

A theory is a collection of associations referring to a specific phenomenon, explained by a system of actions, events, processes and ideas [14]. The objective of a theoretical analysis is to deliver clear views on the causal relationships between certain constructs [7]. The key purpose of using a theory in the analysis is to: (a) describe a certain phenomenon, its causal relationships and its boundaries; (b) clarify how, why and when certain events are occurring; (c) provide approximate forecasts of what will occur under certain conditions; and (d) provide detailed guidance for constructing a certain phenomenon. A conceptual structure to address a particular problem typically needs to be built to help describe a theory [15]. This research seeks to establish a methodological structure focused on a systematic analysis of IoT-related literature and its implementation from a range of perspectives, with specific reference to Saudi Arabia's healthcare services. It promotes the qualitative and quantitative approaches to achieve the study objectives. These are explained in more detail below.

This study's qualitative approach is established in the diffusion of innovation (DOI) concept, which assumes the propagation of new notions (innovations) among those who are within social system is through certain channels as time passes [16]. This study's interview questions were crafted using these factors while the analysis of the interview data was towards ascertaining the most remarkable factor(s) affecting innovation diffusion. The conceptual framework is made up of the beliefs, theories, assumptions as well as anticipations that that supports and informs this study.

Referring to the quantitative method, the goal of this study is to examine the critical factors for IoT adoption in Saudi Arabia from the citizens' perspectives. A conceptual framework – TAM - is devised to achieve the objectives of this analysis. TAM has been chosen for a number of reasons. First, it may provide valuable insights into technology adoption for people using e-government services directly [12]. Second, TAM can be applied flexibly to the IoT adoption in healthcare service studies [12]; [17]; [18]. Third, TAM is robust and accurate in predicting e-government adoption [19]; [20]; [21]. Fourth, TAM confirms that great strides are being made in understanding the use of unique innovations like IoT [22].

III BASIS OF THEORETICAL FRAMEWORKS

This research used DOI theory to explore healthcare administrators' diffusion and adoption of the IoT for delivering healthcare services. DOI theory comprises four key elements, namely, innovation, communication channels, time and social systems [16]. Innovation pertains to the notion, practice or object that a person or other units of adoption see as new [16]. The theory also underscores the importance of perception given that the manner in which users or social systems perceive a given idea determines whether such a notion will be regarded as a noble or desired concept [16]. Thus, it is possible for an idea to have existed for many years and still be viewed as new by people or social systems.

VI DIFFUSION OF INNOVATION THEORY

According to the postulation of the DOI theory, the diffusion process has to do with the movement of new notions (innovations) via some vehicles among the people in a social system [23]. A critical part of this procedure is innovation that comes with the execution of new notions characterised by creating organisational value [24]. New notions can exist as services, systems or procedures or even the improvement of the existing ones. The manner in which new notions (innovations) spread (diffusion) is unique. Researchers need to focus such uniqueness towards the identification of the reasons for quicker proliferations of innovations relative to others.

Compatibility, relative advantage, trialability, complexity and observability directly influence the acceptance of innovations by individuals; that is, an innovation is regarded as superior to a mere idea. Relative advantage is measured on the basis of the social prestige, convenience and satisfaction derived from an innovation, apart from the economic advantages offered by it. Compatibility refers to the status of consistency with existing values and previous experiences and requirements of possible adopters, while complexity has to do with the extent to which an innovation is perceived as easy to comprehend and utilise. Trialability refers to the degree to which an innovation can be subjected to experimentation on a limited basis, and observability is the extent of the visibility of an innovation's outcomes to others.

Studies related to organisational analysis often apply DOI theory in describing the basic patterns in adopting IT innovations [25]; [26]. Rogers [16] who introduced DOI theory looked into innovation adoption as the decision of taking and using new creations. His observation shows that innovation diffusion is strongly dependent on human capital and that there is almost a normal distribution of the number of people who adopt an invention with time.

Wamba [27] carried out a review that suggests DOI theory as a well-known approach in studying IoT adoption. Although it is used extensively in organisational analysis, there have been criticisms regarding this viewpoint due to its individualistic approach in describing organisational behaviours, its inattention to the impacts of environmental factors and its inability to consider aggregate IT implementation [28], [29]. However, some researchers are still supporting this theory. For example the theory was applied in the study of Wamba and Chatfield [30] towards the investigation of the development of value from RFID supply chain projects in logistics and manufacturing domains. Similarly, Quetti, et al. [31] used the theory to examine the RFID adoption procedure in a vertical supply chain with silk industry in Italy being the case study. Liu, et al. [32] also used the theory in the development of a research model intended towards identifying crucial determining factors for the purpose of adopting RFID to deliver public services.

The French sociologist Gabriel Tarde was the first to carry out a study on the diffusion of innovation in 1903 [33]. His research involved the plotting of an S-shaped diffusion curve, which continues to be used today given that the adoption of the majority of new ideas exhibit an S-shaped pattern [33]. Depending on the speed of adoption, the curve's slope can be steep (fast adoption) or gradually inclined (slow adoption) [16]. The S-curve formulated by Tarde was validated by sociologists Bryce Ryan and Neal Gross [34] in their scrutiny of the diffusion of hybrid corn seeds among Iowa farmers in the United States. The cumulative plotting that the authors performed across time demonstrated that the rate at which the hybrid seeds are adopted follows a typical S-shaped curve. Their study motivated further research on the diffusion of innovation (Ryan & Gross, 1943People belonging to the late majority see innovation adoption as an economic necessity and embrace such technologies mostly because of peer pressure [16]. In this group are individuals who exercise caution during the adoption of a new idea [16]. In the laggard group, people do not forgo conventional practices and procedures, and they are unwilling to adopt novel conceptions [16].

The decision to adopt and use an innovation is preceded by the stages of awareness, interest, evaluation and trial [23]. During the awareness stage, people or social systems become conscious of an innovation but lack information or knowledge concerning it. The interest stage is typified by attentiveness, but this phase is still wanting in any active endeavour to search for further information about the innovation. The evaluation stage involves a mental or systemic assessment of the innovation in order to pinpoint how the invention will benefit its current or future users and deciding on whether to embrace it or not. The trial phase is where the innovation is used in its entirety. Adopters make the decision to use an innovation continually [23], but an important barrier to such espousal is uncertainty, which is brought about by the effects of innovations [35].

There is sufficient corroboration that innovations such as the IoT can help improve the healthcare sector [36]. However, the successful implementation of innovations in a particular region does not necessarily mean that such advances will proliferate rapidly to other locations, if at all. Berwick [36] identified three factors that slow down innovation diffusion in healthcare: perceptions regarding innovations, the characteristics of likely adopters and organisational, managerial and contextual factors [36].

Sauer and Lau [37] declared that adopting any process or product engendered through an information system (IS) necessitates decision-making. Decision-makers are in charge of the adoption of new IS innovations because they have the resources and power/authority to drive people to implement behavioural changes in an organisation. In the healthcare industry, the chief decision-makers are healthcare administrators who are also responsible for the adoption of IoT in healthcare services. This reality underscores the need to cast light on the lived experiences of such personnel regarding how they use IoT technologies as part of their service provision responsibilities to patients. DOI theory is generally believed to be a valid model of innovation diffusion, especially in the healthcare domain.

V PROPOSED CONCEPTUAL FRAMEWORK

Given these issues, the present work carried out a qualitative phenomenological analysis to fully elucidate the factors that characterise the slow adoption of the IoT by healthcare organisations in Saudi Arabia. Specifically, the lived experiences of healthcare providers in their use of IoT technologies as they provide healthcare services is explored in detail [38]. The study site is Jazan, which is located in the south-western region of Saudi Arabia, and the study population consists of administrative personnel who make decisions regarding technology adoption. As stated earlier, the theoretical framework that guided the research was DOI theory. The conceptual framework and its constituent elements are illustrated in Fig. 1.

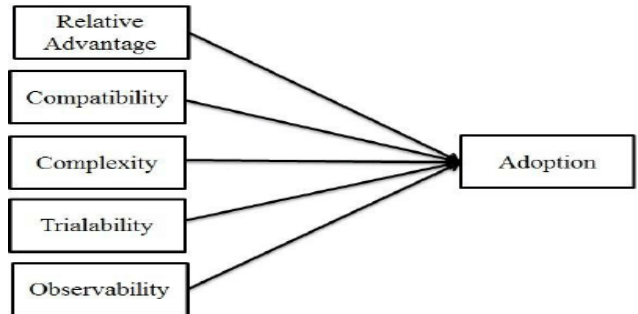

*Figure 1: Diffusion of innovation theory (Rogers, 1983)*

For the quantitative approach, TAM is broadly acknowledged as a great theory for forecasting the technologies' adoption from an individual viewpoint [39]. It is widely used to explicate why people choose to use a particular technology or not. Two main constructs, namely perceived user-friendliness (PEOU) and perceived user-friendliness (PU) are proposed as the determinants influencing individuals' decision to adopt a technology [40]. PEOU refers to the degree to which a person thinks the use of a technology is easy and requires little effort [40]. PU is related to the degree to which a person believes he can improve his or her performance at work using a technology [40]. Overall, TAM focuses on the beliefs of individuals about adopting a technology [40]. The conceptual structure for this study is shown in Figure 3.1 below, as it is an integral part of discussing the crucial factors for IoT adoption in Saudi Arabia. This research covers four types of TAM extension variables. The four categories suggested concentrate on entity, programme, architecture and context aspects. Individual characteristics include computer productivity, personal creativity and computer anxiety. The system features include quality of services and quality of knowledge. The interface features include PEOU and PU.

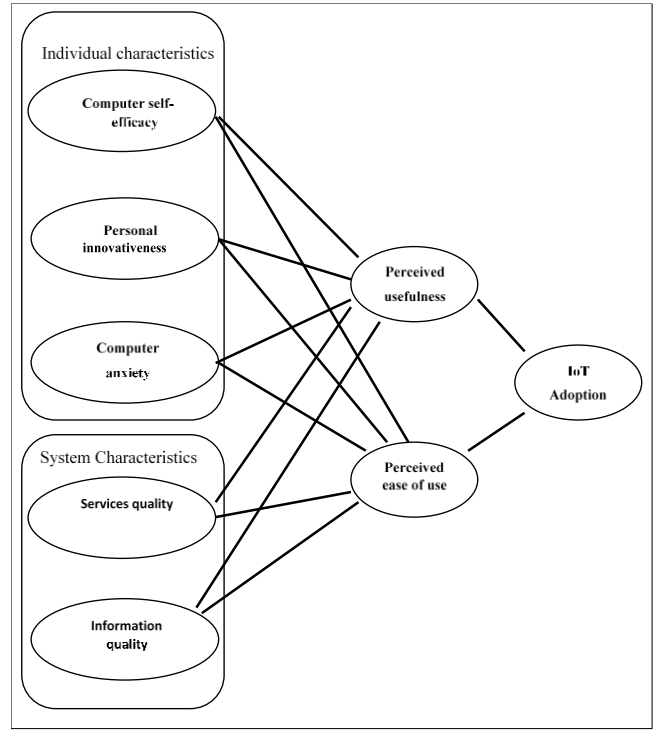

*Figure 3: A Conceptual Framework for the Adoption of IoT*

## VI DATA COLLECTION AND ANALYSIS

This study is planned to collect data using mixed methods, with implementation via semi-structured interviews and survey data. There was the combination of inquiries and procedures while data were obtained from the participants. The analysis of the results was inductive, moving from particular to general issues, and the interpretation of the collected data. In the first phase, the choice of a qualitative method was due to the necessity of exploring a particular phenomenon based on the views and experiences of people.

This research aims to develop an IoT integration framework for healthcare services in Saudi Arabi's e-government development sector. A mixed methodology is used to achieve these research goals. A conceptual framework is developed by assuming the critical factors for effective compliance of IoT integration with healthcare service. The conceptual framework will be tested and validated using structural equation modelling with the use of survey data collected from healthcare administrators in Saudi Arabia. For further validation of the identified critical factors, thematic analysis is conducted on the simultaneously collected semi-structured interview data. The quantitative findings and qualitative findings are triangulated to better understand compliance with healthcare in e-government development of IoT integration in the Kingdom.

Interview is among the most critical data source and proof with the aim of confirming people's perspectives, notions and observations. Many consider an interview as the most remarkable tool to gather deep information about social actors' attitudes, behaviours, observations, knowledge and notions in modern situations [41]. There are three categories under an interview; structured, semi-structured and unstructured designs [42]; [41]. The semi-structured is considered as the most useful and effective in collecting qualitative perceptions, examining and understanding human behaviours [43]; [44]; [41]. The general means of accomplishing this is through open-ended questions, as it permits participants' deep discussion of their experiences and behaviours [43]; [44]. It also offers the chances of understanding a context for exploration and correspondingly connects social situations and social actors' attitudes [42]. To carry out interviews, a critical criterion is participants' agreement to reveal appropriate information regarding their experiences [42]. It is also possible to explore more themes as well as appropriate information [45].

Questionnaires, which require participants to respond to an identical set of questions, are important data collection tools [41] as they enable a researcher to identify the variability of a given phenomenon. Although it is possible to use questionnaires as the only data collection instrument in a study, combining it with other methods is recommended under a mixed methods design [42]. Questionnaires are one of the most popular data collection instruments in research on business and education. Many people have experience using questionnaires [41]. For the current study, the questionnaire method was regarded as the best way to acquire relevant quantitative data in the second phase of the investigation.

For this paper, the use of a quantitative study approach is suitable for two purposes. First, by collecting and analysing numerical data, a quantitative methodology may explain the fundamental link between the research constructions [46]; [47]. Secondly, the implementation of such a technique enables the generalisation of study results for large populations [48]; [49]. This study adopts the survey approach using the quantitative data collection technique. It is widely used to investigate the cause of the phenomenon and collect data on sample population attitudes and behaviour [50]; [49]. This research adopts a survey on the basis of its capacity to study (a) the existing trends of IoT adoption in the Kingdom and (b) testing and validating special hypotheses for the framework suggested in this study.

All the data collected in the qualitative stage were examined using thematic content analysis with the help of NVivo software. The data analysis included the examination, organisation, categorisation and interpretation of the data derived from qualitative and quantitative sources [51]. No standardised method or technique has been developed for the analysis of qualitative data [52]. Hence, the most useful analytical approach is selected to achieve the objectives of a study. The term 'qualitative' is often used as a synonym for 'interview'. It refers to the use of non-numerical data, thus pertaining typically to informational forms other than words, such as pictures or video clips [52]. The basic purposes of qualitative research are the examination of a research topic in accordance with the views and opinions of participants to understand why and how people develop and maintain certain points of view about a situation or problem [53].

During the COVID 19, ensuring that the study does not violate any social isolation and community lockdown for maintaining human spacing, communication and interactions, we redefine new context for conducting the study and set up some variables if they are applicable. We consider the situation of COVID 19 and will make sure that the research will be sensitive to avoid direct human contact. Also, the research will ensure that the appropriate social isolation and the community lockdown process does not violate Saudi and Australia instructions. Additionally, the research will be sensitively maintained. In this case, the ethics application might be considered to be changed.

VII DISCUSSION AND CONCLUSION

The proposed methodological approach was designed to capture both the issues that emerged as important themes in the literature review as well as the results of the year one thesis. Responses from the health administrator to the questions illustrate areas that need more study. In other words, utilising the methodological lens described in this paper, the outcomes would give an adequate insights in terms of answering the research questions. Given that the first year results provided potential interviewees for this study, they may be asked to extend and further interpret of the initial responses. The proposed methodology would assist the Saudi government to develop strategies for developing further on research investigation for useful insights generation leading to implementing IoT in healthcare based on the comments elicited by participants in the target community.

The overarching research question is for investigation through the proposed research approach is what are the factors that influence the effective integration of IoT innovations and healthcare service delivery in Saudi Arabia? This investigation is expected to add new and important information about the integration of IoT innovations and healthcare service delivery in Saudi Arabia. IoT is now considered to be advantageous in various industries including agriculture, hospitality, advertising, construction, manufacturing, services and healthcare. Integrating IoT with the existing technical system would revolutionise Saudi Arabia's healthcare/medical system. Since the Saudi government expects to provide the best possible healthcare to its citizens, IoT integration is essential. Integration of IoT in the technical aspects of healthcare will widen the scope of treatment, link devices with people, applications and machines, and help provide better care for patients, make processes more efficient, reduce errors and save much time and avoid delays. The aim of this research is to give a methodological overview to develop a real problem-based scenario for developing an innovative healthcare information system (for example [10], and [11], in future. This research provides an opportunity for future research to develop a new design framework (e.g. [54] and [55]) that influence the adoption of IoT in any form of new healthcare information system design for instance [56] or [57].